\documentstyle[preprint,aps,psfig]{revtex}
\newcommand{\rar}{\rightarrow}
\tightenlines
\begin{document}
\pagestyle{empty}
\preprint{ICN-UNAM 97-06}
\title{$H_2^+$ ion in strong magnetic field: a variational study}
\author{J.C. Lopez\thanks{E-mail:  vieyra@xochitl.nuclecu.unam.mx}\ 
	P. Hess\thanks{ E-mail:  hess@xochitl.nuclecu.unam.mx} and 
	A. Turbiner\thanks{On leave of absence from the Institute for
	Theoretical and Experimental Physics, Moscow 117259,
	Russia. \hbox{E-mail: turbiner@axcrnb.cern.ch, 
	turbiner@xochitl.nuclecu.unam.mx}}}
\address{Instituto de Ciencias Nucleares, UNAM, Apartado Postal 70-543,
	04510 M\'exico D.F., M\'exico}
\date{\today}

\maketitle

\begin{abstract}
Using a single trial function we perform an accurate calculation of
the ground state $1\sigma_g$ of the hydrogenic molecular ion $H^+_2$
in a constant uniform magnetic field ranging $0-10^{13}$ G.  We show
that this trial function also makes it possible to study the negative
parity ground state $1\sigma_u$.  We obtain that over the whole range
of magnetic fields studied, the calculated binding energies are in
most cases larger than binding energies obtained previously by other
authors using different methods.

\end{abstract}


\narrowtext

\pagestyle{plain}

\section{Introduction}
For a long time the behavior of atomic and molecular systems in a
strong magnetic field of order $B\geq 10^9$ G has attracted
considerable attention.  The main interest comes from both
astrophysics -- studies of white dwarfs and neutron stars (see
\cite{Ruderman:1974,Khersonskii:1984}, and also, for example
\cite{Lai:1995,Lai:1996}, and the review \cite{Liberman:1995} and
references therein), as well as from chemistry -- formation of unusual
chemical compounds whose existence is impossible without strong
magnetic fields \cite{Kadomtsev:1971} (for a review, see, for example,
\cite{Schmelcher:1995} and references therein).

There are many studies of the hydrogen atom $H$ -- the simplest atomic
system -- in a strong magnetic field, while the hydrogen molecular ion
$H^+_2$ -- the simplest molecular system which is stable with respect
to dissociation $H^+_2 \rar H + p$ -- is much less explored. One of
the major drawbacks of many of these studies is a restricted domain of
applicability: they are accurate in the weak magnetic field region but
are inappropriate for the strong magnetic field region and vice
versa. The goal of the present Note is to carry out an accurate
variational calculation of $H^+_2$ in  magnetic fields ranging from
0 up to $10^{13}$ G
\footnote{where the relativistic corrections can still be neglected
(see a discussion in \cite{Liberman:1995} and references therein)}
using a {\it unique} simple trial function equally applicable for any
value of the magnetic field strength. We restrict our consideration to
the case  where the magnetic field is directed along the axis of the
molecule, which is evidently the optimal configuration leading to the
lowest energy.  Our main perception is that the calculations should
not be technically complicated and also  easily reproduced, while the
trial function should be simple enough to allow  further analytic and
numerical investigations.

\narrowtext
\section{Choice of trial functions} 
A constructive criterion for an adequate choice of trial function was
formulated in \cite{Turbiner:1980} and further development was
presented in \cite{Turbiner:1984,Turbiner:1987}. In the simplest form
the criterion is the following. The trial function $\Psi_t(x)$ should
contain all symmetry properties of the problem in hand. If the ground
state is studied, the trial function should not vanish inside of the
domain where the problem is defined. The potential
$V_t(x)=\frac{{\mathbf\nabla}^2 \Psi_t}{\Psi_t}$, for which the trial
function is an eigenfunction, should reproduce the original potential
near singularities and also its asymptotic behavior. The use of this
simplest possible recipe has led to a unique one-parameter trial
function, which, in particular made it possible to carry out the first
qualitative study of the ground state of the hydrogen molecule $H_2$
in the region of both weak and strong magnetic fields
\cite{Turbiner:1983}. Later a few-parameter trial function was
proposed for a description of the hydrogen atom in an arbitrary
magnetic field, which led, for the low-excited states, to an accuracy
comparable with the best calculations
\cite{Turbiner:1987,Turbiner:1989}.

Now we wish to apply the above-mentioned recipe to the ion $H_2^+$.
We work in the Born-Oppenheimer approximation. Let us first introduce
notation (see Fig.1).  We consider two attractive centers of charge
$Z$ situated on the $z$-axis at a distance $R/2$ symmetrically with
respect to the origin.  The magnetic field of the strength $B$ is
directed along the $z$ axis and $r_{1,2}$ are the distances from the
electron to the first(second) center, respectively. The quantity
$\rho$ is the distance from the electron to the $z$-axis. Through the
paper the Rydberg is used as the energy unit. For the other quantities
standard atomic units are used. The potential corresponding to the
problem we study is given by
\begin{equation}
\label{e1}
V\ =\ \frac{2Z^2}{R} -\frac{2Z}{r_1} - \frac{2Z}{r_2} + \frac{B^2
\rho^2}{4} \ ,
\end{equation}
where the first term has the meaning of the classical Coulomb energy
of the interaction of two charged centers.

One of the simplest functions satisfying the above recipe is the
Heitler-London function multiplied by the lowest Landau orbital:
\begin{equation}
\label{e2}
\Psi_1= {e}^{-\alpha_1 Z (r_1 + r_2) - \beta_1 B \rho^2 /4},
\end{equation}
where $\alpha_1,\beta_1$ are variational parameters. It has in total 3
variational parameters if we include the internuclear distance $R$, in
the search for the equilibrium distance.  It is known that in the absence
of a magnetic field a function of the Heitler-London type gives an
adequate description of diatomic systems near their equilibrium
position. The potential corresponding to this function is:
\begin{eqnarray}
\label{e3}
V_1 &=& 2 Z^2\alpha^2_1 -B \beta_1 -2 Z\alpha_1
\Big(\frac{1}{r_1}+\frac{1}{r_2}\Big)+\frac{\beta^2_1 B^2\rho^2}{4} \nonumber\\
&{}&+ \ 2Z^2\frac{\alpha^2_1}{r_1r_2}\Big[\rho^2 +(z-R/2)(z+R/2)\Big]
+
Z\alpha_1 B \beta_1 \rho^2 \Big(\frac{1}{r_1}+\frac{1}{r_2}\Big) \ .
\end{eqnarray}

It is clear that this potential reproduces the original potential (1) near
Coulomb singularities and at large distances, $|x| \rightarrow \infty$.

The Hund-Mulliken function multiplied by the lowest Landau orbital is another
possible trial function:
\begin{equation}
\label{e4}
\Psi_2= \Big( {e}^{-\alpha_2 Z r_1} + e^{-\alpha_2 Z r_2} \Big)
{e}^{-\beta_2 B \rho^2 /4} \ ,
\end{equation}
where $\alpha_2,\beta_2$ are variational parameters. It is well known
that this function, in the absence of a magnetic field, describes the
region of large internuclear distances. The calculations we performed
show that this property remains valid for all magnetic fields up to
$10^{13}$ G.  Like (2), the trial function (4) is characterized by 3
variational parameters.  In order to take into account both
equilibrium and large distances, we should use an interpolation of (2)
and (4).  There are two natural ways to interpolate:
\begin{itemize}
\item[(i)] a non-linear superposition:
\begin{equation}
\label{e5}
\Psi_{3-{1}}= \Big( {e}^{-\alpha_3 Z r_1 -\alpha_4 Z r_2} + \sigma
{e}^{-\alpha_4 Z r_1 -\alpha_3 Z r_2 } \Big)
{e}^{-\beta_3 B\rho^2/4}  \ ,
\end{equation}
where $\alpha_{3}, \alpha_{4}, \beta_3$ are variational
parameters. The parameter $\sigma=\pm 1$ depends on whether positive
parity $1\sigma_g$ or negative parity $1\sigma_u$ states we consider.
The function (5) is a modification of the Guillemin-Zener function
used for the description of the molecular ion $H^+_2$.  If
$\alpha_{3}= \alpha_{4}$, the function (5) reduces to (2) and if
$\alpha_{3}=0$, it coincides with (4). In total there are 4
variational parameters characterizing the trial function (5);
\item[(ii)] a linear superposition of (2), (4)
\begin{equation}
\label{e6}
\Psi_{3-{2}}= A_1 \Psi_1 + A_2 \Psi_2  \ ,
\end{equation}
\end{itemize}
where the relative weight of (2) and  (4) in (6) is taken as an extra variational
parameter. It is a 7-parameter trial function.

Of course, as a natural continuation of the above interpolation procedure
one can take a linear superposition of all three functions:
the modified Heitler-London, Hund-Mulliken and Guillemin-Zener functions
(2), (4), (5)
\begin{equation}
\label{e7}
\Psi_4 = A_{3-{1}}\Psi_{3-{1}}+ A_{3-2}\Psi_{3-{2}} =
 A_{3-{1}}\Psi_{3-{1}} + A_1 \Psi_1 + A_2 \Psi_2 \ ,
\end{equation}
where again, as in the case of the function (6) the relative weights
of different components are variational parameters. In total, the
trial function (7) is characterized by 10 variational parameters.
Most of our calculations will be carried out using this function. The
minimization procedure was done using the  standard minimization
package MINUIT from CERN-LIB on a Pentium-Pro PC.

\narrowtext
\section{Results}
In order to present our results we begin with the study of the
dependence of our variational results on different trial functions
(see Table 1).  It turned out that all variational parameters are of
the order of one independently on the value of the magnetic field
strength.  In the Table 2 we give a comparison of our calculations
with the best known results. Since we are doing a variational study of
the problem, the (binding) energies obtained represent (lower) upper
bounds to the exact energies. No need to mention that many
calculations were performed for the case of the $H^+_2$ ion in absence
of a magnetic field. Our results are in agreement with the best
calculations within an absolute accuracy of $10^{-5}$. For all studied
values of the magnetic field ($\le 10^{13}$ G) our results for binding
energies exceed the best known results at the present.  Among all
previously made calculations, we should emphasize that those performed
by Wille \cite{Wille:1988} have the most extended domain of
applicability. The accuracy of these results is almost as good as the
accuracy of our results for magnetic fields $B\le 10^{11}$ G. However
for bigger magnetic fields the accuracy of his results fells down
drastically.

 Fig.2 shows the electronic density distribution as a function of
magnetic field.  For a small magnetic field the distribution has two
clear maxima corresponding to the positions of the centers. The
situation changes drastically for magnetic fields of the order of
$B\simeq 3 \times 10^{11}$ G, where the probability of finding the
electron in any position between the two centers is practically the
same. For larger magnetic fields the electron is preferably located
between the two centers with the maximum in the middle, $z=0$ (see Fig
2 (e), (f)). Due to a loss of accuracy this phenomenon was not
observed in \cite{Wille:1988}. It is worth noticing that for all
magnetic fields studied the region of large internuclear distances is
dominated by the Hund-Mulliken function (4).  

In Table 3 the results of the study of the ground state of negative
parity $(1\sigma_u)$ are presented. Let us note that in the absence of
a magnetic field the electron is not bounded possessing a shallow
minimum (see, for example, \cite{Teller:1930,Peek:1980}).  However, in
a magnetic field the electron becomes bounded in agreement with
general expectations \cite{Kadomtsev:1971} (see also \cite{Peek:1980}).

In conclusion we want to emphasize that we present the most accurate
calculations for the ground state energies and equilibrium distances
of the molecular ion $H^+_2$ in magnetic field. Unlike the majority of
other studies our calculations stem from a unique framework covering
both weak and strong magnetic field regimes.

\acknowledgements
The authors wish to thank M. Ryan for reading of the manuscript and  comments.

This work is supported in part by DGAPA project  IN105296

\newpage

\widetext
\begin{figure}[h]
\centerline{\hbox{
\psfig{figure=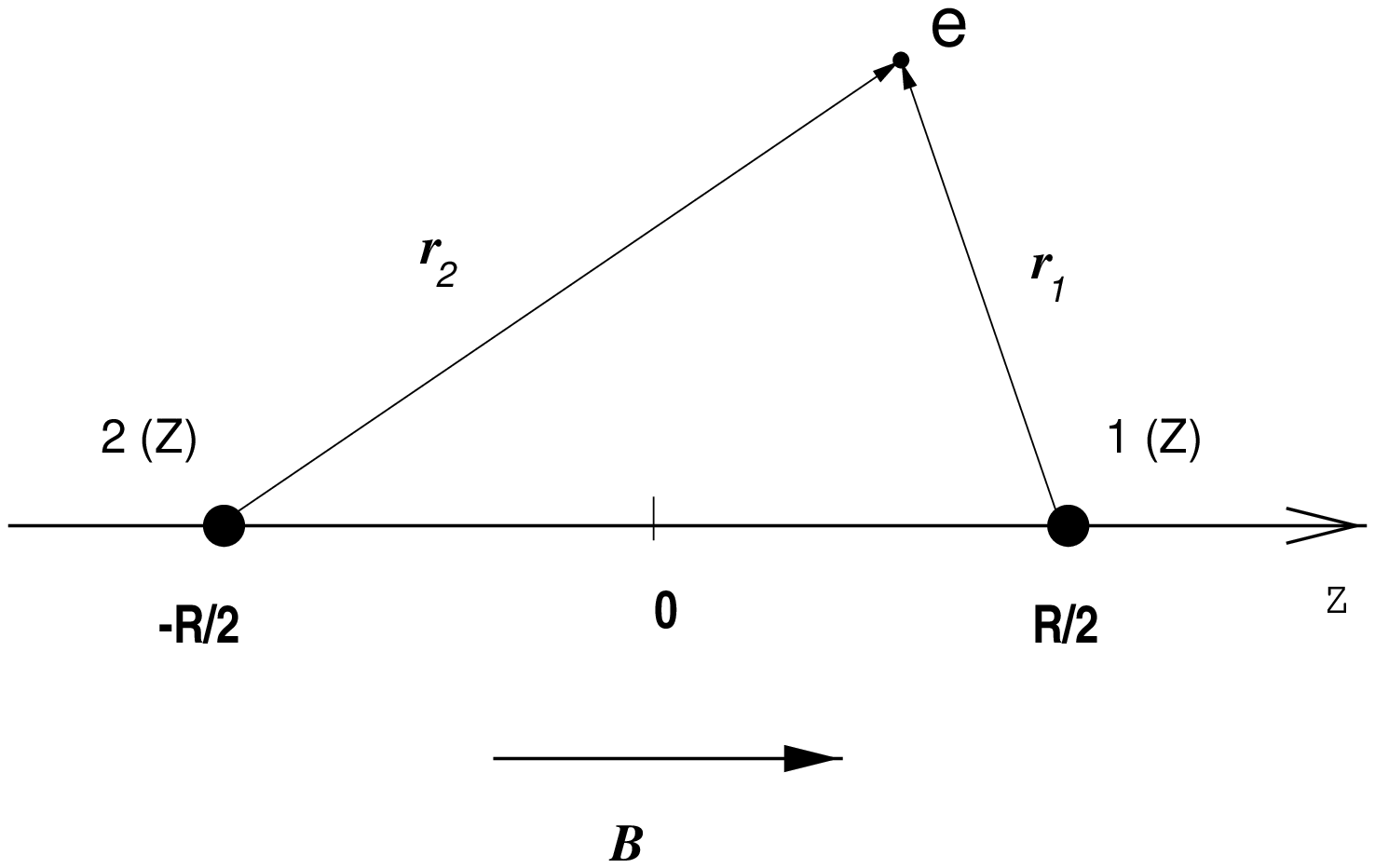,width=5.5in,angle=-90}
}}
\caption{$H_2^+$ in a magnetic field $B$. Explanation of the notations used}
\end{figure}

\newpage

\widetext
\begin{figure}
\centerline{\hbox{
\psfig{figure=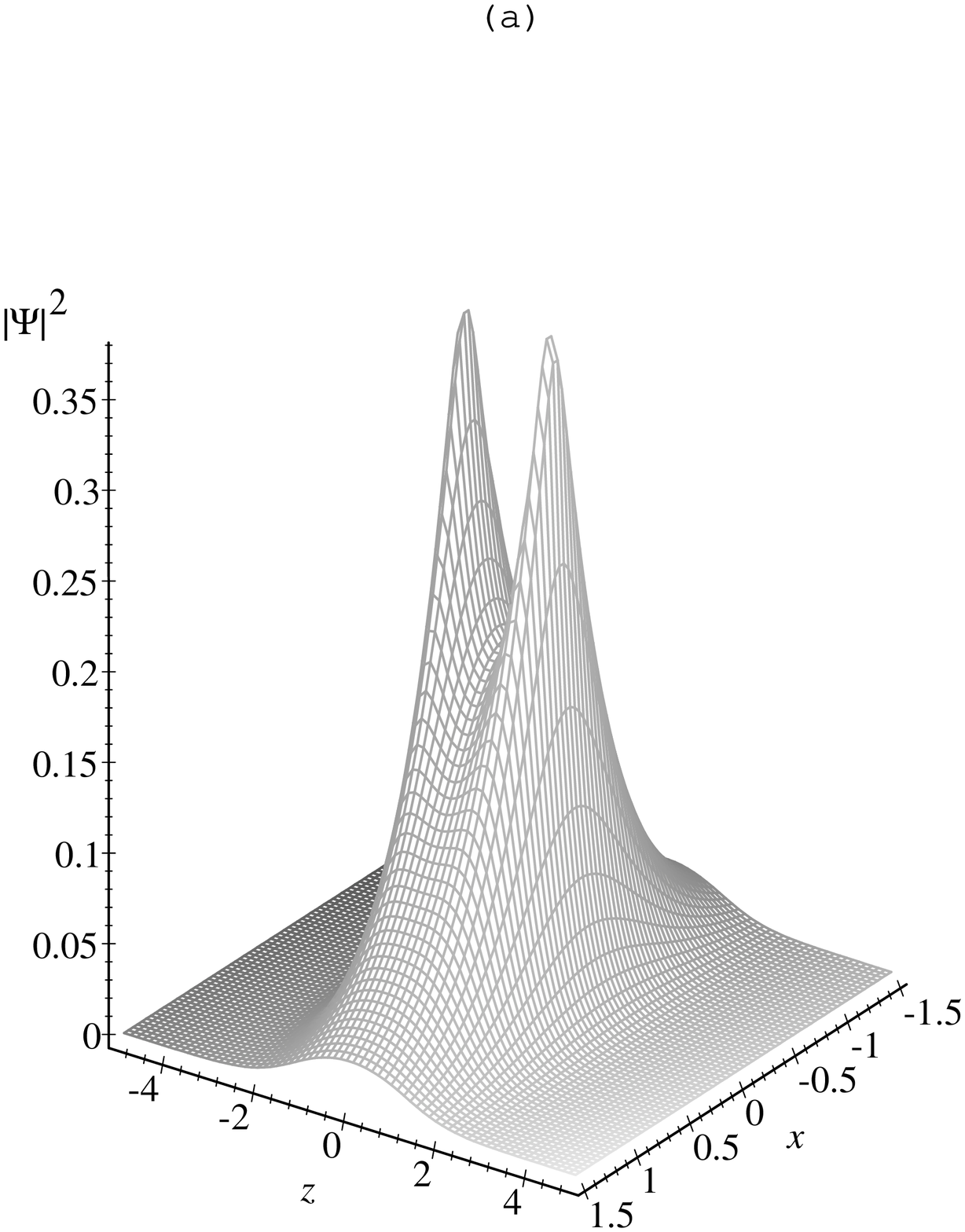,width=2.8in,height=3.2in}
\psfig{figure=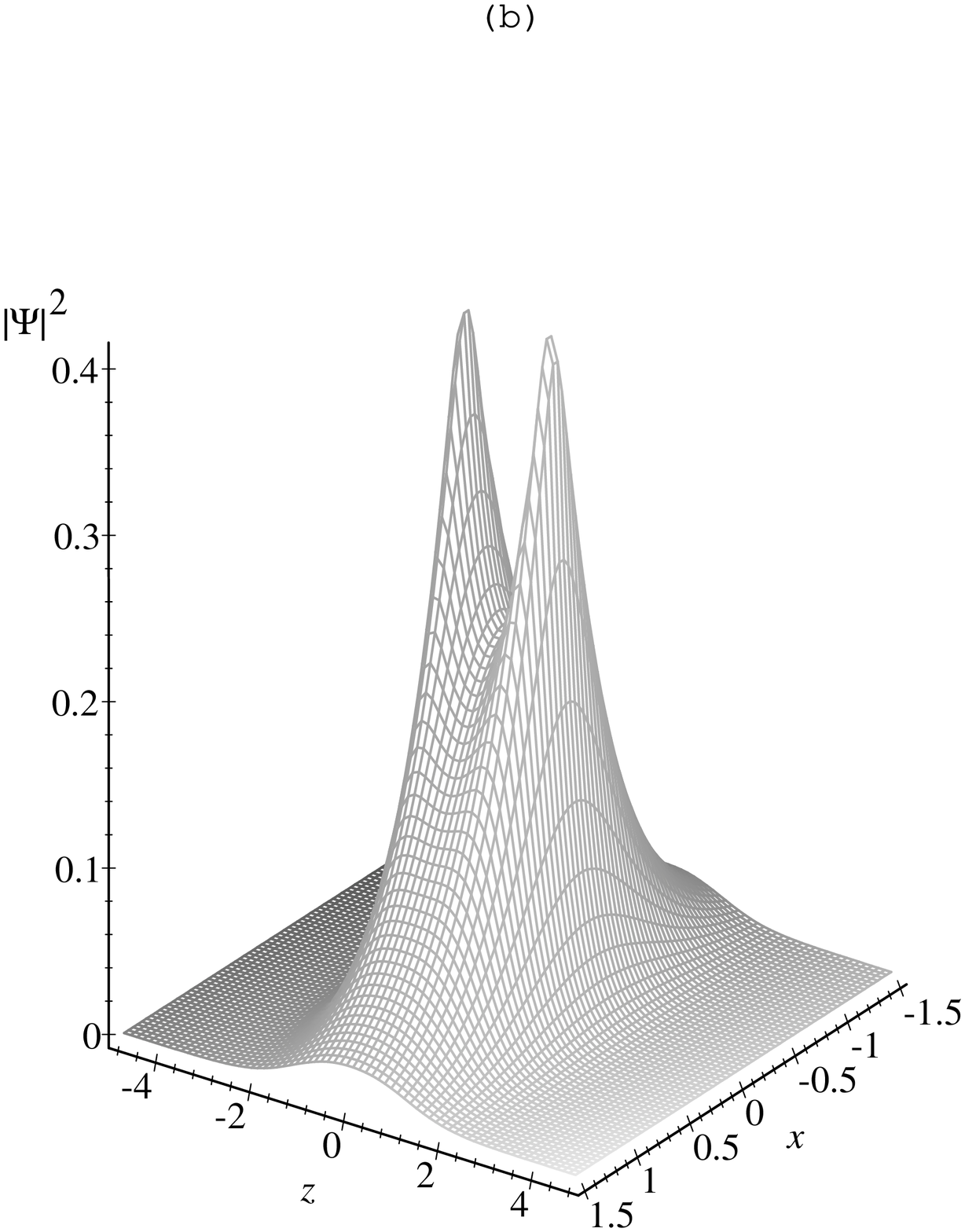,width=2.8in,height=3.2in}
}}

\centerline{\hbox{
\psfig{figure=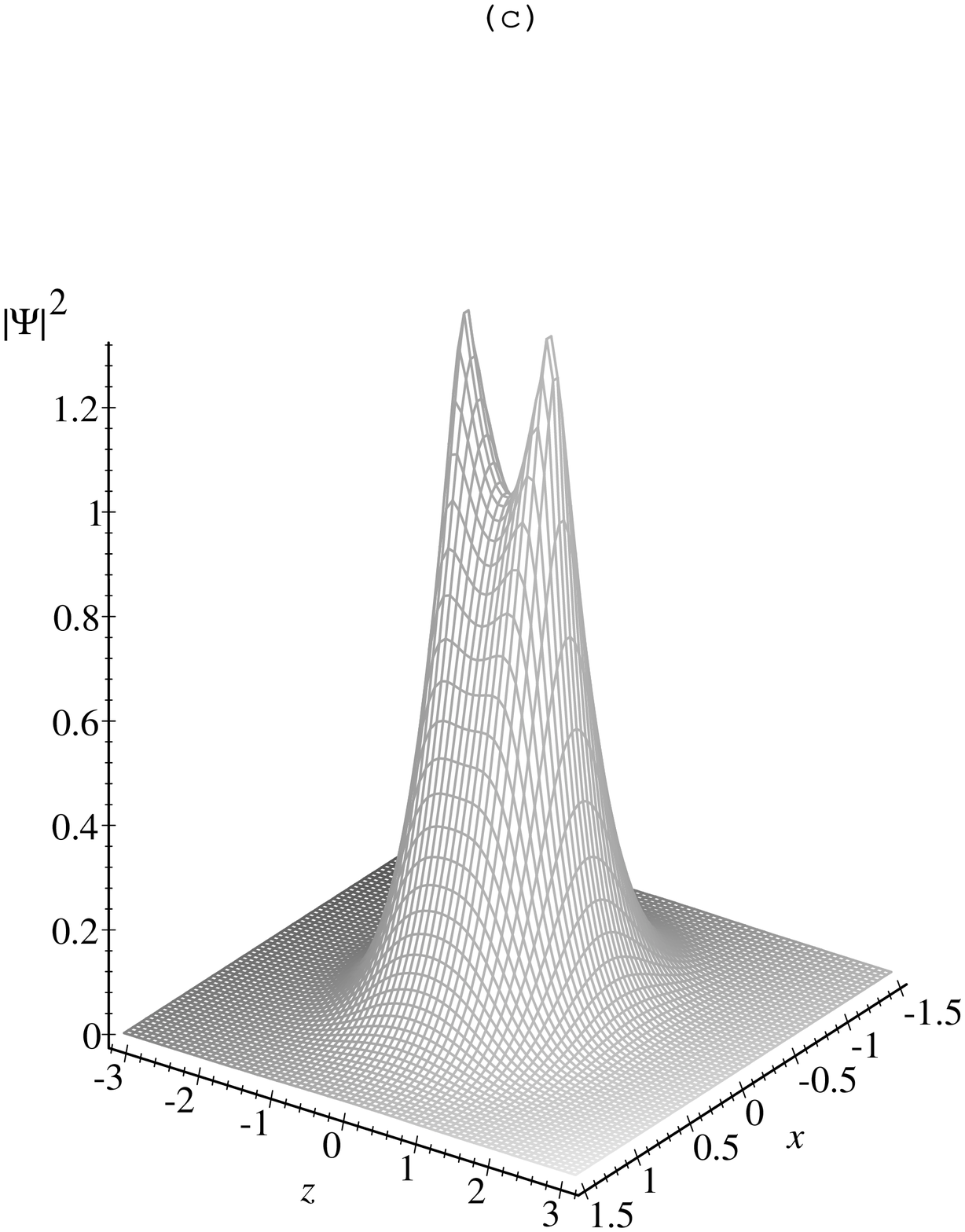,width=2.8in,height=3.2in}
\psfig{figure=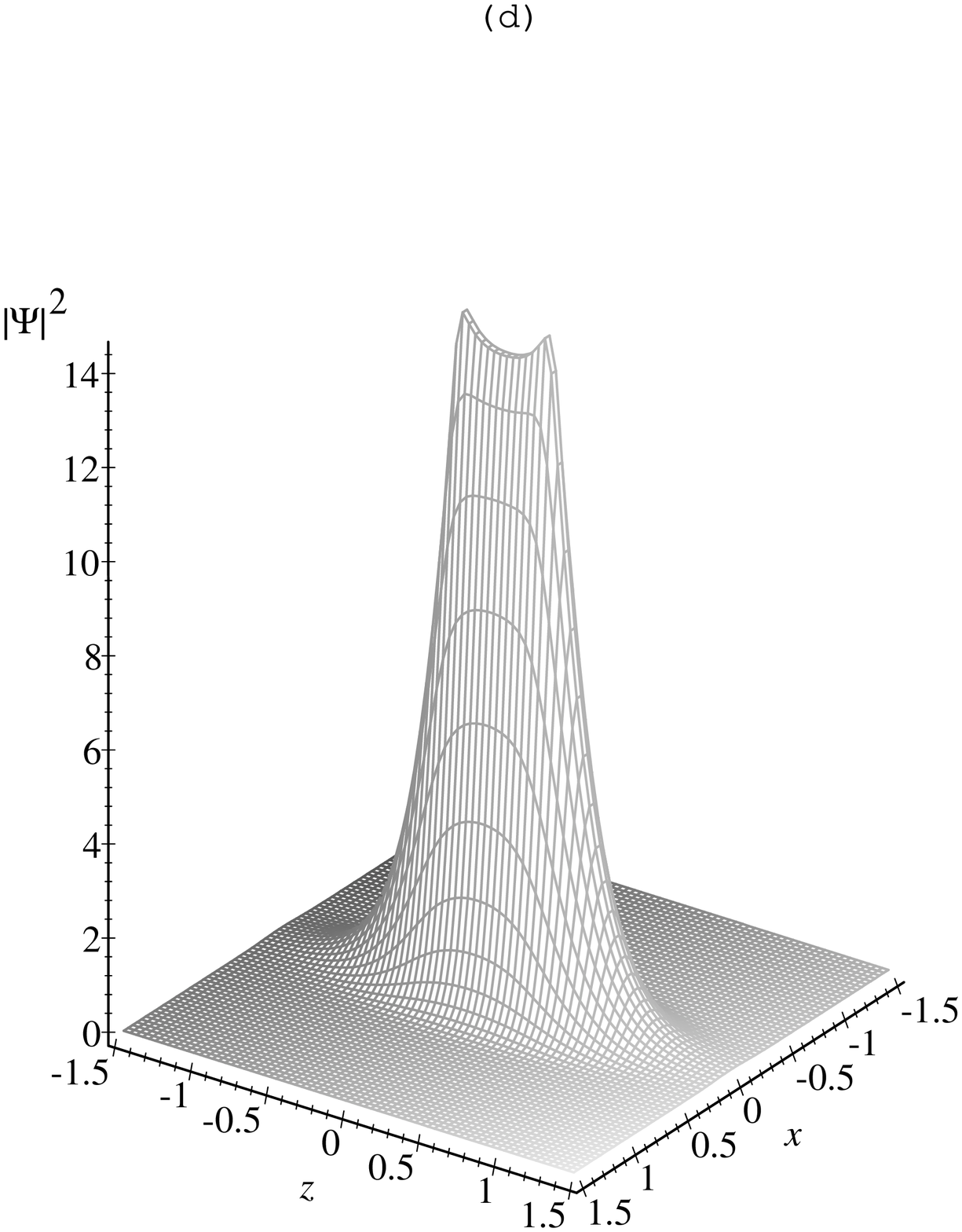,width=2.8in,height=3.2in}
}}
\caption{The electronic probability density of the ground state
$1\sigma_g$ at $y=0$ as a function of the magnetic field strength. (a)
$B=0$, (b) $B=10^9$ G, (c) $B=10^{10}$ G, (d) $B=10^{11}$ G.}
\end{figure}

\addtocounter{figure}{-1}

\widetext
\begin{figure}
\centerline{\hbox{
\psfig{figure=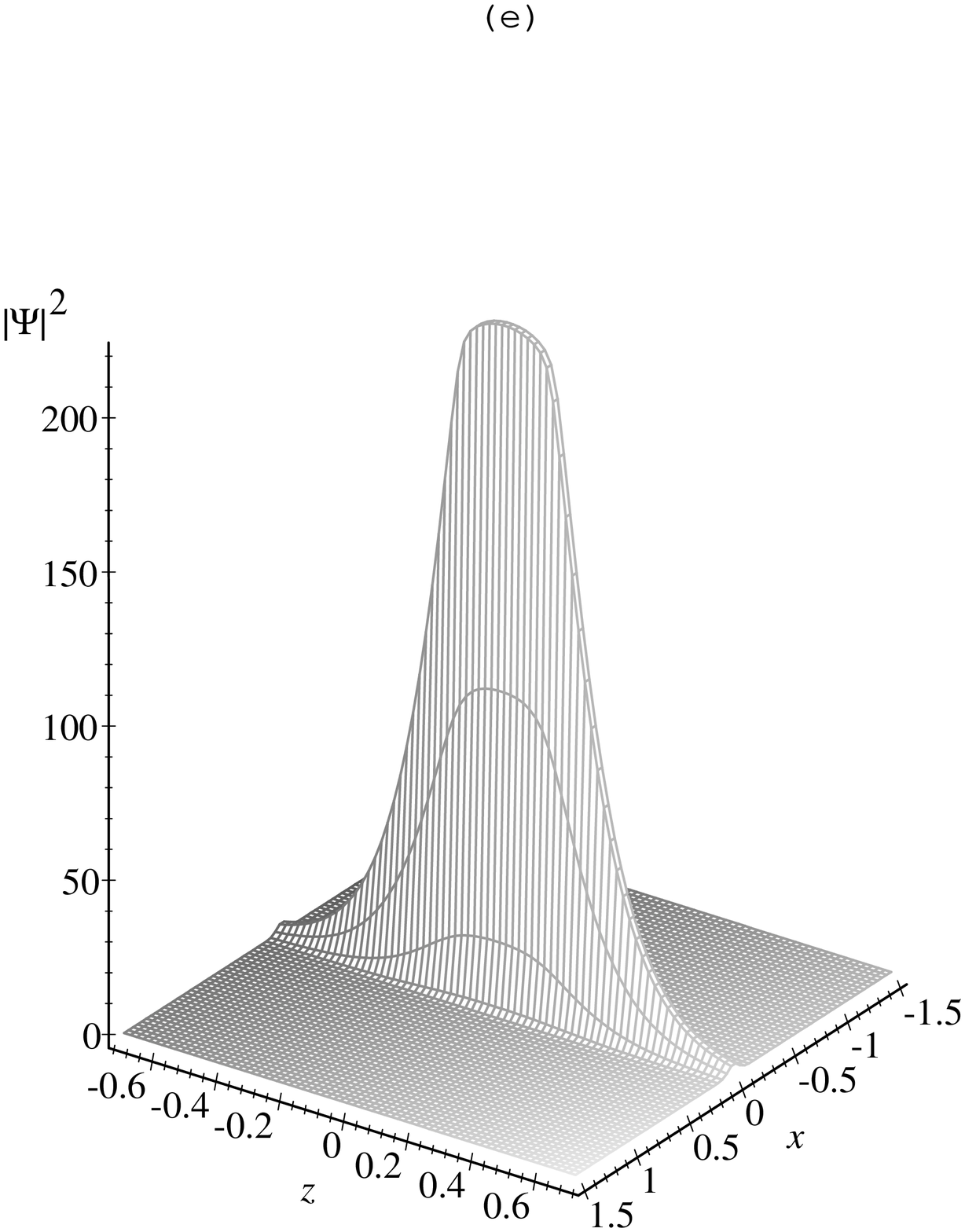,width=2.8in,height=3.2in}
\psfig{figure=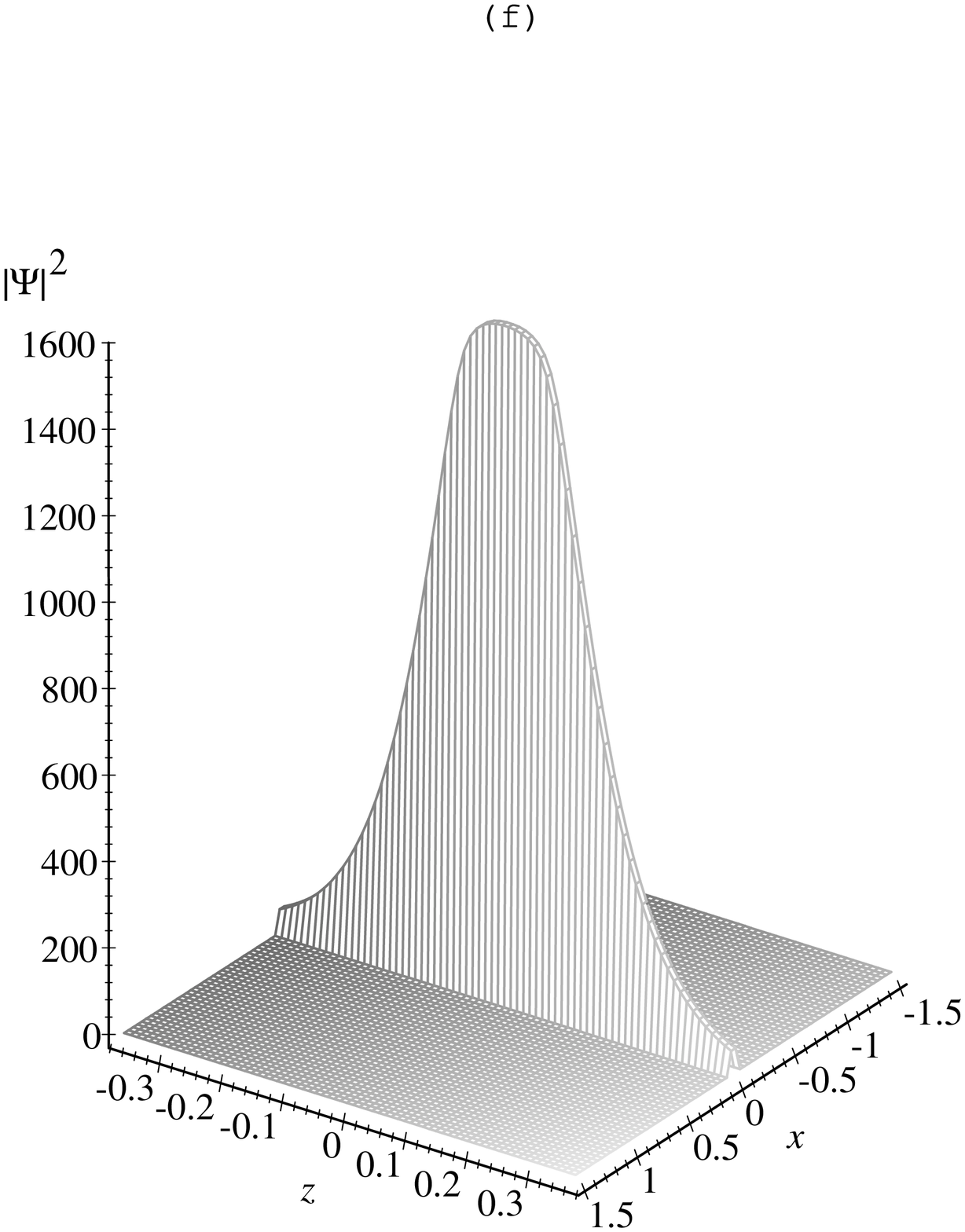,width=2.8in,height=3.2in}
}}
\caption{(Continuation) (e) $B=10^{12}$ G, (f) $B=10^{13}$ G.}
\end{figure}

\widetext
\begin{table}
\caption{Ground state energies and the equilibrium distances for the
ground state $1\sigma_g$ of the ion $H_2^+$ obtained using different trial
wave functions as a function of the magnetic field strength.}

\begin{tabular}{cccccccc}
$B$        & & $\Psi_1$    & $\Psi_2$   &$\Psi_{3-1}$ & $\Psi_{3-2}$ & $\Psi_4$\\ \hline 
           &    &          &         &         &         &          \\
$0$        
           &$E $& -1.16277 &-1.17299 &-1.20488 &-1.20488 &-1.20525   \\
           &$R $& 1.84678  & 2.00349 & 1.99799 & 1.99810 & 1.99706\\ 
           &    &          &         &         &         &          \\
           &    &          &         &         &         &          \\
$10^9$ G    
           &$E $&-1.11760 &-1.11849 &-1.15007 &-1.15013 &-1.15071   \\
           &$R $& 1.80967 & 1.93353 & 1.92378 & 1.92375 & 1.92333\\ 
           &    &          &         &         &         &          \\
           &    &          &         &         &         &          \\
$10^{10}$ G             
           &$E $& 1.11835 & 1.15713 & 1.10098 & 1.10095 & 1.08993   \\
           &$R $& 1.21434 & 1.24328 & 1.25042 & 1.25013 & 1.24640   \\ 
           &    &          &         &         &         &          \\
           &    &          &         &         &         &          \\
$10^{11}$ G             
           &$E $& 35.1065 & 35.2351 & 35.1016 & 35.0750 &35.0374     \\
           &$R $&  0.59575&  0.59206&  0.60065&  0.58954& 0.59203   \\ 
           &    &          &         &         &         &          \\
           &    &          &         &         &         &          \\
$10^{12}$ G 
            &$E $&408.539  &408.837  &408.539  & 408.424 &408.300    \\
            &$R $&0.28928  &0.28372  &0.28927  & 0.28422 & 0.28333\\
            &    &          &         &         &         &          \\
\end{tabular}
\end{table}

\newpage

\widetext
\begin{table}
\caption{Comparison of present calculations for the ground state 
${1\sigma_g}$ of equilibrium distance ${R}$, energy ${E}$ and binding energy
${BE = B-E}$ with the results of other calculations [4,13-17].}
\begin{tabular}{lcccl}
 $B$ (Gauss) & $R$ (a.u.)& $E$ (a.u.)     &  $BE$ (a.u.)      & \ \ \
           \ Source        \\ \hline 
           &         &          &             &                   \\
$B=0$      
           & 1.9971  & -1.20525 &    ---      & Present           \\
           & 2.0000  & -1.20527 &    ---      & Teller \cite{Teller:1930} \\ 
           & 1.997   & -1.20527 &    ---      & Wille \cite{Wille:1988} \\
           & 1.997   & -1.20526 &    ---      & Peek--Katriel \cite{Peek:1980}      \\
           & 2.0000  & -1.205268&    ---      & Wind \cite{Wind:1965} \\
           &         &          &             &                   \\ 
$B=10^9$ G 
           & 1.9233  & 1.15072  &  1.57616    & Present   \\    
           & 1.924   & 1.15072  &  1.57616    &  Wille \cite{Wille:1988} \\
           & 1.921   &   ---    &  1.5757     & Peek-Katriel \cite{Peek:1980}\\
           & 1.90    &  ---     &  1.5529     & Lai--Suen\cite{Lai-Suen:1982} \\
           &         &          &             &                   \\ 
$B=10^{10}$ G 
           & 1.2464  & 1.08979  &  3.1646     & Present       \\
           & 1.246   & 1.09031  &  3.1641     & Wille  \cite{Wille:1988} \\
           & 1.159   &  ---     &  3.0036     & Peek--Katriel \cite{Peek:1980}\\
           & 1.10    &  ---     &  3.0411     & Lai--Suen\cite{Lai-Suen:1982} \\
           &         &          &             &                   \\


$B=10^{11}$ G 
	   & 0.593   & 35.0362  &  7.5080     & Present       \\
           & 0.593   & 35.0428  &  7.5013     & Wille \cite{Wille:1988}\\
           & 0.62    &  ---     &  7.35       &Lai--Salpeter\cite{Lai:1996} \\ 
           &         &          &             &                   \\
$B=5\times 10^{11}$ G
	   & 0.351   & 199.238  & 13.483      & Present       \\
	   & 0.350   & 199.315  & 13.406      &  Wille \cite{Wille:1988}     \\
           & 0.35    &  ---     & 13.38       & Lai--Salpeter\cite{Lai:1996} \\ 
           &         &          &             &                   \\
$B=10^{12}$ G 
	   & 0.283   & 408.300  & 17.141      & Present       \\
	   & 0.278   & 408.566  & 16.875      &  Wille \cite{Wille:1988}         \\
           & 0.28    &  ---     & 17.06       & Lai--Salpeter\cite{Lai:1996} \\ 
           &         &          &             &                   \\
$B=2\times 10^{12}$ G
           & 0.230   & 829.274  & 21.609      & Present \       \\
           & 0.23    &  ---     & 21.54       & Lai--Salpeter\cite{Lai:1996} \\
           &         &          &             &                   \\ 
$B=5\times 10^{12}$  G           
	   & 0.177   & 2098.3   & 28.954      & Present       \\
           & 0.18    &  ---     & 28.90       &Lai--Salpeter\cite{Lai:1996} \\
           &         &          &             &                   \\
$B=10^{13}$ G           
	   & 0.147   & 4218.7   & 35.752      & Present       \\
           & 0.15    &  ---     & 35.74       & Lai--Salpeter \cite{Lai:1996} \\
           &         &          &             &                   \\
\end{tabular}

\end{table}

\newpage

\widetext
\begin{table}
\caption{Energy, binding energy and equilibrium distance for the ground state
of $H_2^+$ of negative parity $1\sigma_u$. }
\begin{tabular}{lcccl}
 $B$ (Gauss) & $R$ (a.u.)& $E$ (a.u.)     &  $BE$ (a.u.)  & \ \ \ \
Source  \\ \hline
           &         &          &             &                  \\
$B=0$           
	   &  12.746 & -1.00010 &  1.00010    &   Present         \\ 
           &  12.55  & -1.00012 &   1.00012   &   Peek-Katriel\cite{Peek:1980}\\ 
           &         &          &             &                  \\
$B=10^9$ G
           & 11.039  & -0.92063 &  1.34608    &  Present          \\ 
           & 10.55   & -0.917134&  ---        &  Peek-Katriel\cite{Peek:1980}\\ 
           &         &          &             &                  \\
$B=10^{10}$ G
           & 6.4035  & 1.6585   &  2.59592    &  Present          \\ 
           & 4.18    & 2.1294   &   ---       &  Peek-Katriel\cite{Peek:1980}\\ 
           &         &          &             &                  \\
$B=10^{11}$ G 
           & 3.7391  & 36.945   & 5.59901     &  Present          \\ 
           &         &          &             &                  \\
$B=10^{12}$ G
           & 2.4329  & 413.92   & 11.519      &  Present          \\ 
           &         &          &             &                  \\
$B=10^{13}$ G
           & 1.7532  & 4232.6   & 21.851      &  Present          \\
           &         &          &             &                  \\ 
\end{tabular}
\end{table}

\end{document}